\documentclass[onecolumn,letterpaper,12pt,journal]{IEEEtran} 
\usepackage{graphicx}
\usepackage{color}
\usepackage{float}
\usepackage{amssymb}
\usepackage{amsthm}
\usepackage{cite}
\usepackage{amsmath}
\usepackage{algorithm}
\usepackage{mathrsfs}

\usepackage[leftmargin=0cm,rightmargin=0cm,innerleftmargin=0cm,innerrightmargin=0cm]{mdframed}
\usepackage[letterpaper, left=.7in, right=.7in, top=.78in, bottom =.78in]{geometry}
\usepackage[noend]{algpseudocode}

\usepackage[labelformat=simple]{subcaption}

\newtheorem{definition}{Definition}

\newmdenv[topline=false,bottomline=false]{leftrightbot}

\newcommand{\mat}[1]{\boldsymbol{\mathrm{#1}}}
\newcommand{\set}[1]{\mathcal{#1}}
\newcommand{\sset}[1]{\boldsymbol{\mathcal{#1}}}

\DeclareMathOperator*{\argmax}{arg\,max}

\newcommand{\br}[1]{{\left\{ #1 \right\}}}
\newcommand{\abs}[1]{{\left| #1 \right|}}

\begin{document}

\title{Distributed Clustering for Multiuser Networks through Coalition Formation}
\author{Rami~Mochaourab, Eduard~Jorswieck, Mats Bengtsson
\thanks{This work has been submitted to the IEEE for possible publication. Copyright may be transferred without notice, after which this version may no longer be accessible.}
\thanks{Rami Mochaourab and Mats Bengtsson are with School of Electrical Engineering, KTH Royal Institute of Technology, 100 44 Stockholm, Sweden. E-mail: \{rami.mochaourab, mats.bengtsson\}@ee.kth.se. 
Eduard Jorswieck is with Communications Theory, Communications Lab, TU Dresden, 01062 Dresden, Germany. E-mail: eduard.jorswieck@tu-dresden.de.}}


\maketitle

%
The vast increase in the number of mobile communication devices necessitates new technological advances in mobile broadband networks. Due to the natural limitations on the communication and computational resources, it is essential to develop suitable signal processing and resource allocation mechanisms which are scalable with the number of devices. First, the scarcity of the communication resources, such as bandwidth and power, demands efficient resource sharing schemes that take into account the interdependence in the devices' performance measures \cite{Luo2008}. Second, the large computational overhead associated with centralized system-wide optimization in large networks can be impractical such that distributed mechanisms become the proper alternative \cite{Bacci2016,Bayat2016}. 


%
%
Consider, as an illustrative example, the multiple-input single-output interference channel (MISO~IC)~\cite{Liu2011}, in which multiple transmitter-receiver pairs (links) operate on the same spectral band. Each transmitter uses~$t > 1$ antennas while each receiver has a single antenna. Then, the received signal at a receiver~$i$ is given~by
%
\begin{equation}\label{eq:example}
	y_i = \underbrace{\mat{h}_{ii}^\dag \mat{w}_i s_i}_\text{desired} + \underbrace{\sum\nolimits_{j\neq i} \mat{h}_{ji}^\dag \mat{w}_j s_j}_\text{interference} + \underbrace{n_i}_\text{noise}, 
\end{equation}
\noindent where $\mat{h}_{ji} \in \mathbb{C}^{t}$ is the channel vector between transmitter $j$ and receiver $i$, $\mat{w}_{i} \in \mathbb{C}^{t}$ is the beamforming vector used by transmitter $i$, $s_i \sim \mathcal{CN}(0,1)$ is the signal intended for receiver $i$, and $n_i\sim \mathcal{CN}(0,\sigma^2)$ is additive white Gaussian noise. From \eqref{eq:example}, it is evident that interference will be the main cause for network inefficiency, and hence must be managed through cooperative beamforming. Since optimal cooperative beamforming, e.g., to maximize the links' sum rate, is generally difficult to determine in the MISO IC \cite{Liu2011}, we must resort to suboptimal but efficient beamforming strategies. In addition, in order to support distributed algorithms we may restrict the beamforming schemes to rely only on local information at the transmitters. Depending on the designed transmission schemes, which will directly affect the links' performance, cooperation in the network may become beneficial only within clusters of links. As a result, the network optimization problem boils down to finding efficient clustering in which each link only cooperates with the members of its own cluster. Such problems are related to coalition formation games which provide us with basic tools for developing distributed clustering algorithms.

\section{Relevance}

Clustering mechanisms are essential in certain multiuser networks for achieving efficient resource utilization. The main purpose of this lecture is to present the theory of coalition formation which is useful for distributed clustering problems. We reveal the generality of the theory, which is currently missing in the literature, and study complexity aspects which must be considered in multiuser networks.


\section{Prerequisites}
Basic knowledge in set theory is required.
\section{Problem Formulation}

\begin{figure*}[t]
  \centering
  \begin{minipage}[t]{5.3cm}
  \centering
  \includegraphics[width=5.3cm,clip]{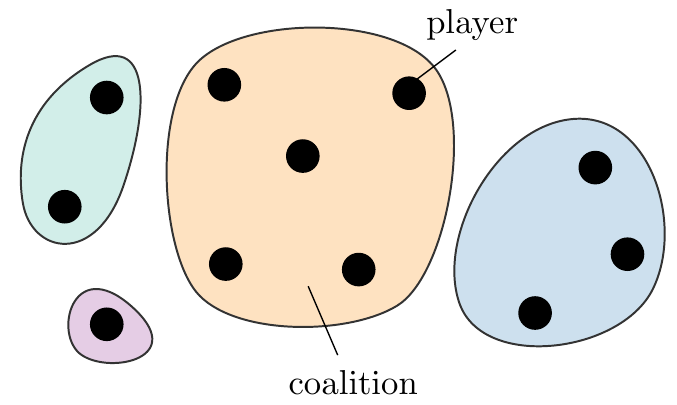}%
  \subcaption{\label{fig:coalition_structure} Coalition structure $\sset{C}$}
  \end{minipage}
  \hspace{1cm}
  \begin{minipage}[t]{5.3cm}
  \centering
  \includegraphics[width=5.3cm,clip]{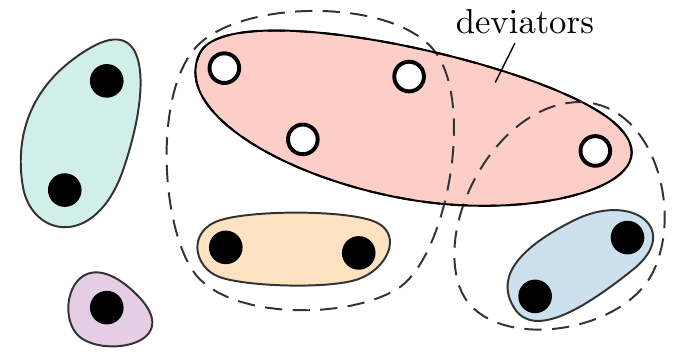}%
  \subcaption{\label{fig:group_deviation} General deviation}
  \end{minipage}

  \begin{minipage}[t]{5.3cm}
  \includegraphics[width=5cm,clip]{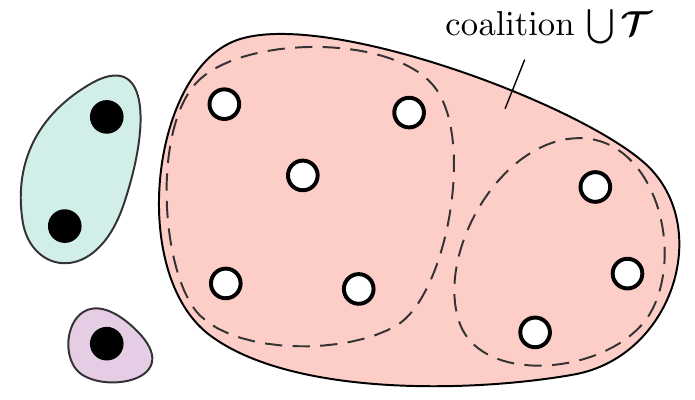}
  \subcaption{\label{fig:merge} Merge}
  \end{minipage}%
  \hfill
  \begin{minipage}[t]{5.3cm}
  \includegraphics[width=5cm,clip]{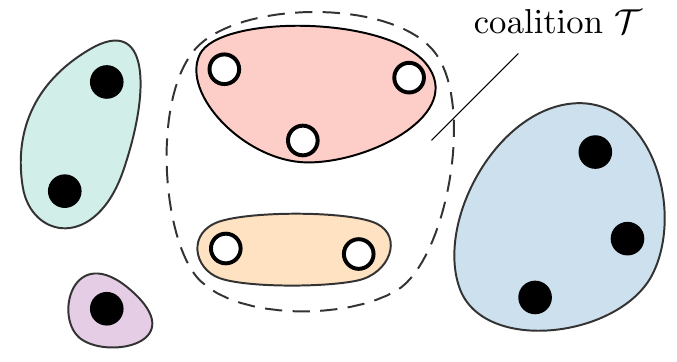}
  \subcaption{\label{fig:split} Split}
  \end{minipage}%
  \hfill
  \begin{minipage}[t]{5.3cm}
  \includegraphics[width=5cm,clip]{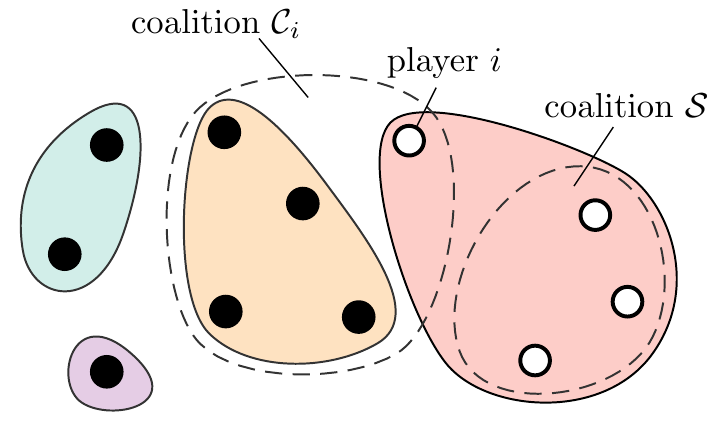}
  \subcaption{\label{fig:deviation} Individual deviation}
  \end{minipage}
  \vspace{-.2cm}
  \caption{Illustrations for the deviation models.}
  \vspace{-.6cm}
\end{figure*}

Let $\set{N}$ be the set\footnote{Throughout, we will use calligraphic font for sets and boldface calligraphic font for sets of sets.} of devices in the network, e.g., the set of links in the MISO IC. A clustering of the network, as is illustrated in \figurename~\ref{fig:coalition_structure}, will be called a \emph{coalition structure} \cite{Thrall1963}.
\begin{definition}[Coalition structure]\label{def:coalition_structure}
A {coalition structure} $\sset{C}$ is a partition of $\set{N}$ into a set of pairwise disjoint coalitions. Let $\sset{P}$ be the set containing all coalition structures. 
\end{definition}%
Each element of a coalition structure $\sset{C}$ is called a \emph{coalition} and comprises a set of cooperating devices. Let $\set{C}_i \in \sset{C}$ denote the coalition which includes device $i\in \set{N}$ in coalition structure~$\sset{C}$. For suitable application in communication networks, the devices' cooperation strategies within a coalition must be designed taking into account the available local information along with the communication constraints between the devices. For the MISO IC, we assume that each transmitter has perfect local channel state information (CSI), i.e., each transmitter $i$ knows $\mat{h}_{ij}$ for all $j$ perfectly. With this information and assuming a total power constraint $p_i$, one conventional cooperative beamforming strategy for transmitter $i$ nulls the interference at the receivers in its coalition $\set{C}_i$ while maximizing the desired signal power:
\begin{equation}\label{eq:zeroforcing}
\mat{w}_{i}(\set{C}_i) = \argmax_{||\mat{w}_i||^2 \leq p_i} ~ |\mat{h}_{ii}^\dag \mat{w}_i| \text{ s.t. } \mat{h}_{ij}^\dag \mat{w}_i = 0, \text{ for } j\in\set{C}_i\backslash \{i\}.
\end{equation}
The cooperative beamforming strategy in \eqref{eq:zeroforcing} is called \emph{zero forcing beamforming} and admits a simple closed form expression. Observe that \eqref{eq:zeroforcing} also includes the \emph{noncooperation} strategy of a transmitter $i$, namely when $\set{C}_i = \{i\}$, which corresponds to maximum ratio transmission $\mat{w}_{i}(\{i\}) = \sqrt{p_i} |\mat{h}_{ii}|/||\mat{h}_{ii}||$.


The devices' performance measures, called \emph{utility functions}, are functions of the devices' strategies. In the MISO IC example, we model the utility function of link $i$ as its achievable rate
\begin{equation}\label{eq:utility}
u_i(\sset{C}) = \log_2 \Bigg( 1 + \frac{ |\mat{h}_{ii}^\dag \mat{w}_i(\set{C}_i)|^2}{\sum\nolimits_{\substack{j \in \set{N}\setminus\set{C}_i}} |\mat{h}_{ji}^\dag \mat{w}_j(\set{C}_j)|^2 + \underbrace{\sum\nolimits_{\substack{k \in \set{C}_i\setminus\{i\}}} |\mat{h}_{ki}^\dag \mat{w}_k(\set{C}_k)|^2}_{=0} + \sigma^2}\Bigg),
\end{equation}
\noindent which depends on the beamforming strategy in~\eqref{eq:zeroforcing}. Here we have assumed a block flat-fading channel model and single-user decoding at the receivers. Observe, that the utility function in \eqref{eq:utility} depends on the overall coalition structure $\sset{C}$ through the selected cooperation strategy in \eqref{eq:zeroforcing}. Then, the effect of cooperation within coalition $\set{C}_i$ is evident through interference-nulling by the transmitters in the same~coalition.

We will assume that the devices are able to make local decisions and adapt their transmission strategies in a way to increase their associated utility functions. Relating to the MISO IC example, a transmitter~$i$ would seek to cooperate with a set of links in $\set{C}_i$ using~\eqref{eq:zeroforcing} only if this improves its achievable rate  in~\eqref{eq:utility}. Such modeled behavior for the devices conforms with the game theoretic assumption of players' rationality, and is a main characteristic for supporting distributed mechanisms in multiuser networks \cite{Bacci2016}. To this end, we will refer to the devices as \emph{players}. 


On characterizing the players in the network and modeling their utility functions depending on the coalition structure, the clustering problem can be formulated as a \emph{coalitional game in partition form} \cite{Thrall1963} 
\begin{equation}\label{eq:coalitiona_game}
(\set{N}, \mat{u}).
\end{equation}
\noindent In \eqref{eq:coalitiona_game}, $\set{N}$ is the set of players and $\mat{u}:\sset{P} \rightarrow \mathbb{R}^{\abs{\set{N}}}$ is called the \emph{partition function} which assigns to each player a utility depending on the coalition structure, e.g., the utility function in~\eqref{eq:utility}. The solution of \eqref{eq:coalitiona_game} is a set of coalition structures which are stable according to a certain stability criterion. The problem of finding stable coalition structures is the study of coalition formation games, which we consider next. 

\section{Coalition Formation}
%
While the coalitional game in partition form in \eqref{eq:coalitiona_game} characterizes the conflict between the players through their utility functions, a \emph{coalition formation game} specifies the dynamics that change the coalition structures to reach a solution of \eqref{eq:coalitiona_game}. A coalition formation game is formally represented with the pair \cite{Diamantoudi2007} 
\begin{equation}
(\sset{P},\gg), 
\end{equation}
where $\sset{P}$ is the set of all coalition structures (Definition~\ref{def:coalition_structure}) and $\gg$ is a binary relation on $\sset{P}$, called the \emph{dominance relation} \cite{Diamantoudi2007}. The dominance relation $\gg$ compares two coalition structures in $\sset{P}$ and will be defined as the intersection of two other binary relations in the following.

\begin{definition}[Dominance relation]\label{def:dominance}
Coalition structure $\sset{C}'$ dominates $\sset{C}$, written as $\sset{C}' \gg \sset{C}$, if and only if (iff) $\sset{C}' \succ \sset{C}$, indicating that $\sset{C}'$ is preferred to $\sset{C}$, and $\sset{C}' \xleftarrow[]{} \sset{C}$, indicating that $\sset{C}'$ is reachable from $\sset{C}$.
\end{definition}
The subsequent two sections will deal with the modeling of the dominance relation defined above. Here, we briefly reveal the purpose of the two binary relations which define it: The preference relation $\succ$ on $\sset{P}$ will be used to compare two different coalition structures based on the players' utility functions, i.e., the partition function $\mat{u}$ in \eqref{eq:coalitiona_game}. The reachability binary relation $\xleftarrow[]{}$ on $\sset{P}$ will determine whether it is possible to \emph{change} a coalition structure to another coalition structure by a feasible \emph{deviation} of some players, as illustrated in~\figurename~\ref{fig:group_deviation}. The players' deviation models can be distinguished between: 
\begin{itemize}
\item \emph{Group-based deviation}: A \emph{group} of players, possibly members of different coalitions, can leave their coalitions and form a new coalition. 
\item \emph{Individual-based deviation}: A \emph{single} player leaves its coalition to join another. 
\end{itemize}

The reachability binary relation will affect the number of possible ways to change a given coalition structure. In multiuser networks, this change would typically require communication between the deviators, cf. \figurename~\ref{fig:group_deviation}. Therefore, the feasibility of communication between the devices will directly affect the modeling of the ``reachability'' binary relation. We will explore different parameterized deviation models later in the next two sections.

By the assumption of player rationality, a coalition structure $\sset{C}$ will change to $\sset{C}'$ if $\sset{C}'$ dominates $\sset{C}$. The following notion of coalition structure stability is called \emph{deviation-proof} and is motivated by \cite{Apt2009,Diamantoudi2007}.

\begin{definition}[Stability]\label{def:stability}
Coalition structure $\sset{C} \in \sset{P}$ is stable iff there exists no $\sset{C}' \in \sset{P}$ such that $\sset{C}' \gg \sset{C}$.
\end{definition}

A stable coalition structure indicates that there exists no other coalition structure which is reachable from the stable coalition structure and is preferred by the players. While reachability depends on the feasibility of communication between the devices in multiuser networks, using the preferences of the players as a criterion for stability accounts for efficiency in the outcome. Stable coalition structures are in this regard suitable clustering solutions for multiuser networks.

On modeling the dominance relation (Definition~\ref{def:dominance}), coalition formation is realized through Algorithm~\ref{alg:coalition_formation}. The algorithm is initialized at a suitable coalition structure $\sset{C}$. First, in Step~\ref{alg_line:deviation} a coalition structure $\sset{C}'$ is found which is reachable from $\sset{C}$. Then, Step~\ref{alg_line:preference} checks whether $\sset{C}'$ is preferred by the players compared to $\sset{C}$. If so, the coalition structure changes to $\sset{C}'$. This process is repeated until no dominating coalition structures are found, i.e., stability is reached. Note that the convergence of Algorithm~\ref{alg:coalition_formation} must be studied for the specific application which defines the dominance relation. 


\begin{algorithm}[t]
\caption{\label{alg:coalition_formation} Generic coalition formation algorithm.}
\begin{algorithmic}[1]
\State \textbf{Input}: $(\sset{P},\gg)$, coalition structure $\sset{C} \in \sset{P}$
\Repeat
\State Find $\sset{C}'$ such that $\sset{C}' \xleftarrow[]{} \sset{C}$ \label{alg_line:deviation}
\State If $\sset{C}' \succ \sset{C}$ then $\sset{C}$ is updated to $\sset{C}'$ \label{alg_line:preference}
\Until {stability}
\State \textbf{Output}: {$\sset{C}$}
\end{algorithmic}
\end{algorithm}
%
%
%
\section{Group-based Deviation}
Group-based deviation describes the mechanism in which a set of players, possibly in different coalitions, build a single coalition, while the remaining players do not change their coalitions. An example is given in \figurename~\ref{fig:group_deviation}, where the coalition structure $\sset{C}$ in \figurename~\ref{fig:coalition_structure} changes (in a single deviation step) because the set of deviators form a single coalition. This deviation model is considered in \cite{Diamantoudi2007} with the assumption of farsighted players, meaning that the players are able to anticipate the effects of a sequence of deviations. In this lecture note, we rather assume \emph{nearsighted} players, i.e., the players can only know the effects of a single deviation.

A special type of group-based deviation is called \emph{nested deviation} \cite{Apt2009,Diamantoudi2007} which restricts the change in the coalition structure as follows: 1) Several existing coalitions can only merge to form a single coalition. 2) A single existing coalition can split to form smaller coalitions. We will describe parameterized versions of such nested deviations in the following two subsections.


\subsection{Merge Deviation Model}
The merge deviation allows a set of at most $q$ coalitions in some coalition structure $\sset{C}$ to \emph{merge} and form a single coalition \cite{Mochaourab2014}. Accordingly, the coalition structure $\sset{C}$ changes to $\sset{C}'$. This mechanism is illustrated in~\figurename~\ref{fig:merge}, where two coalitions comprising $\sset{T}$ form a single coalition $\bigcup \sset{T}$. This deviation model defines the reachability binary relation $\sset{C}' \xleftarrow[\textrm{merge}]{} \sset{C}$ as follows.
\begin{definition}[$q$-merge]\label{def:coalition_merging}
For all $\sset{C}',\sset{C} \in \sset{P}$, $\sset{C}' \xleftarrow[\textrm{merge}]{} \sset{C}$ is true iff $\sset{C}' = (\bigcup \sset{T}) \cup (\sset{C} \setminus \sset{T})$ with ${\sset{T} \subset \sset{C}}, {\abs{\sset{T}} \leq q}$.
\end{definition}

The parameter $q$ in this model restricts the number of reachable coalition structures and thus affects the complexity of coalition formation. Specifically, the complexity of $q$-merge corresponds to the worst case number of coalition structures $\sset{C}' \in \sset{P}$ for which $\sset{C}' \xleftarrow[\textrm{merge}]{} \sset{C}$ is true. This number directly determines the number of searches in worst case required in Step~\ref{alg_line:deviation} in Algorithm~\ref{alg:coalition_formation}. Given a coalition structure $\sset{C}$, the number of possible ways to merge a set of at least two and at most $q$ coalitions is given by \cite{Mochaourab2014}
\begin{equation}\label{eq:comp_merging}
{D}(\abs{\sset{C}},q) = \sum\nolimits_{j=2}^{\min\{q,\abs{\sset{C}}\}} \binom{\abs{\sset{C}}}{j}.
\end{equation}
Since the number of coalitions in $\sset{C}$ is largest when $\sset{C}$ includes only single-player coalitions, then the worst case complexity is ${D}(\abs{\set{N}},q)$. From Lemma 1 in \cite{Mochaourab2014} we have that for fixed $q < \abs{\set{N}}$, the growth rate of ${D}(\abs{\set{N}},q)$ in $\abs{\set{N}}$ is bounded by a polynomial of degree $q$. While it is desired to set large values for $q$ in order to encompass a large set of reachable coalition structures, it is however important to account for the associated coalition formation complexity which increases exponentially in $q$.

The convergence of Algorithm~\ref{alg:coalition_formation} using the $q$-merge reachability model in Definition~\ref{def:coalition_merging} is guaranteed since only merging of coalitions is allowed and the number of possible merging steps is finite. 

The application of $q$-merge in the MISO IC requires synchronization between the links in order to determine the set of coalitions which deviate. This can be done by consistent enumeration of all \emph{reachable} coalition structures at all links and considering each one in a sequential manner. The consistent enumeration of all reachable coalition structures can be done, e.g., through lexicographic ordering~\cite{Mochaourab2014}. Then, for each reachable coalition structure, the members of the coalitions involved in the deviation send each other a cooperation request. The links' replies to the requests depend on the preference relation, which we discuss later below.


\subsection{Split Deviation Model}
In the split deviation model, a single coalition in some coalition structure $\sset{C}$ can \emph{split} into at most $q$ coalitions. In doing so, the coalition structure $\sset{C}$ changes to $\sset{C}'$ as is illustrated in~\figurename~\ref{fig:split}. 
\begin{definition}[$q$-split]\label{def:coalition_splitting}
For all $\sset{C}',\sset{C}\in\sset{P}$, $\sset{C}' \xleftarrow[\textrm{split}]{} \sset{C}$ is true iff $\sset{C}' = (\sset{C} \setminus \set{T}) \cup \br{\set{S}_1,\ldots,\set{S}_k}$, where $\set{T} \in \sset{C}$, $\set{T} = \bigcup_{j=1}^k \set{S}_j$ and $k \leq q$.
\end{definition}

Analogous to $q$-merge, the convergence of Algorithm~\ref{alg:coalition_formation} with $q$-split is guaranteed since only splitting of coalitions is allowed and the number of splitting steps is finite.

\begin{figure*}[t]
  \centering
  \includegraphics[width=10cm,clip]{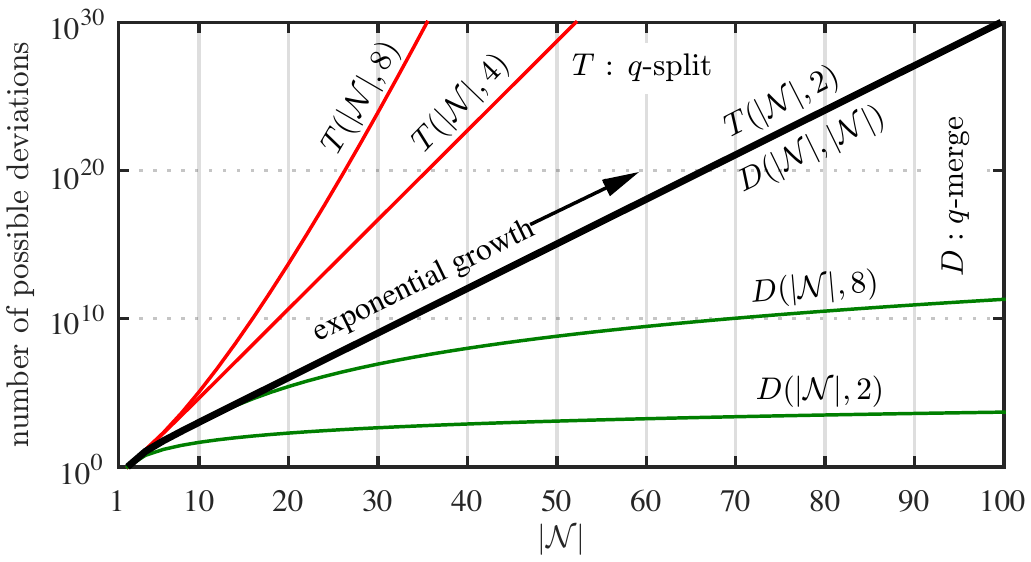}
  \caption{\label{fig:complexity} Complexity of $q$-merge and $q$-split deviations, with $D$ and $T$ defined in \eqref{eq:comp_merging} and \eqref{eq:comp_spliting}, respectively.}
  \vspace{-.6cm}
\end{figure*}
The complexity of the $q$-split deviation model is the worst case number of coalition structures that have to be considered in Step~\ref{alg_line:deviation} in Algorithm~\ref{alg:coalition_formation}. The number of ways to split a set $\set{T}$ into at least two and at most $q$ subsets is \cite{Mochaourab2014}
\begin{equation}\label{eq:comp_spliting}
T(\abs{\set{T}},q) = \sum\nolimits_{\ell=2}^{q} S(\abs{\set{T}},\ell),
\end{equation}
where $S(\abs{\set{T}},\ell) = \frac{1}{\ell!} \sum\nolimits_{k=0}^{\ell}(-1)^{k}\binom{\ell}{k} (\ell - k)^{\abs{\set{T}}}$ are the stirling numbers of the second kind \cite[Ch. 8]{Brualdi2004}. Then, the complexity of Step~\ref{alg_line:deviation} in Algorithm~\ref{alg:coalition_formation} is quantified by $\sum_{\set{T}\in\sset{C}} T(\abs{\set{T}},q)$. For illustration purposes, we consider the complexity of $q$-split for $\sset{C}$ being the grand coalition $\set{N}$ and compare its complexity to $q$-merge for increasing number of players $|\set{N}|$ and different values of $q$ in \figurename~\ref{fig:complexity}. It can be observed that $q$-split has significantly larger complexity compared to $q$-merge; The least complex choice of $q=2$ in the $q$-split deviation model gives an exponential growth ($T(\abs{\set{N}},2)$) which occurs for the most complex $q$-merge deviation model ($D(\abs{\set{N}},\abs{\set{N}})$).

It is possible to define coalition deviation to include a combination of merging and splitting steps~\cite{Apt2009}. The advantages with this deviation model, e.g., allowing more adaptive clustering and guaranteeing efficient clustering outcomes, are discussed in Section ``Coalition Formation Games'' in~\cite{Bacci2016}. Note however that the convergence of coalition formation with sequential merging and splitting is not implied but must be studied depending on the characteristics of the application in multiuser networks.

For the application of $q$-split in the MISO IC example, similar synchronization requirements are needed at the links as in the $q$-merge case, along with a consistent enumeration of all reachable coalition structures.

\subsection{Preference Relations}
A preference relation determines whether a coalition structure $\sset{C}'$ is preferred to another coalition structure $\sset{C}$ based on the utility functions of the players in~\eqref{eq:coalitiona_game}. One preference relation which is appropriate for distributed implementation in multiuser networks, e.g., the MISO IC, is the \emph{Pareto order relation}~$\succ_\set{T}$:
\begin{equation}\label{eq:preference_group}
\sset{C}' \succ_\set{T} \sset{C} \Leftrightarrow u_i(\sset{C}') \geq u_i(\sset{C}) \text{ for all } i \in \set{T},
\text{ and } u_j(\sset{C}') > u_j(\sset{C}) \text{ for some } j \in \set{T}.
\end{equation}
\noindent Note that only a specific set of players $\set{T}$ are considered in $\sset{C}' \succ_\set{T} \sset{C}$, which typically will correspond to the players which deviate. The modeling of the preference relation in~\eqref{eq:preference_group} is suitable for distributed algorithms in multiuser networks since each device in $\set{T}$ knows its utility function and can do the comparison locally. Then, the devices only need to exchange their \emph{binary decisions} to implement the preference relation. 

\section{Individual-based deviation}

In individual-based deviation, a coalition structure can change only if a single player leaves its coalition and joins another \cite{Bogomolnaia2002}. This mechanism is illustrated in \figurename~\ref{fig:deviation}.
\begin{definition}[Individual Deviation]\label{def:deviation}
For all $\sset{C}',\sset{C} \in \sset{P}$, $\sset{C}' \xleftarrow[\textrm{individual}]{} \sset{C}$ is true iff 
\begin{equation}
\sset{C}' = (\sset{C} \setminus \{\set{C}_i,\set{S}\}) \cup \{\set{C}_i\setminus \{i\}, \set{S} \cup \{i\}\},\text{ for some } i \in \set{N} \text{ and } \set{S} \in (\sset{C} \setminus \set{C}_i) \cup \{\}.
\end{equation}
\end{definition}

The convergence of Algorithm~\ref{alg:coalition_formation} with the individual-based deviation model will generally depend on the used preference relation and the utilities of the players~\cite{Bogomolnaia2002}. It is however possible to enforce convergence through suitable restrictions on the deviation model. One widely used approach prevents a player from joining a coalition which it has been a member of before~\cite{Saad2012,Brandt2015a}. According to Theorem 1 in~\cite{Saad2012}, this restriction ensures convergence of the algorithm irrespective of the used preference relation.

In the MISO IC example, individual deviation requires selecting a single link at a time, according to a random or fixed procedure, which asks the members of an existing coalition to join their coalition. If the preference relation allows the deviation to take place, then the link informs the members of its coalition.

%
%
While there exist several preference relations which are suitable for individual-based deviation~\cite{Bogomolnaia2002}, we mention in this lecture note one preference relation which considers both the utilities of the \emph{deviating} player $i$ and the members of the coalition $\set{S}$ it requests to join:
\begin{equation}\label{eq:preference_individual}
\sset{C}' \succ_{i,\set{S}} \sset{C} \Leftrightarrow u_i(\sset{C}') > u_i(\sset{C}), \textrm{ and } u_j(\sset{C}') \geq u_j(\sset{C}) \text{ for all } j \in \set{S}.
\end{equation}
The above preference relation is suitable for application in multiuser networks, because the members of $\set{S}$ are required to cooperate with player $i$, such as by updating their beamforming strategies in the example MISO IC. Then, it is reasonable to demand that the utilities of the players in $\set{S}$ do not decrease.

\section{Discussion}

Coalition formation games essentially provide structured procedures which are useful for solving clustering problems. The significance of coalition formation games are their ability to exploit the users' performance measures, provided these are formulated to depend on the clustering, in order to obtain efficient outcomes. Moreover, owing to the players' rationality assumption, coalition formation games particularly support the design of distributed clustering mechanisms.

The complexity of coalition formation mainly depends on the used deviation model, which ultimately impacts the overhead in communication between the players and the overhead in cooperation within a coalition. While conceptually different, nested group-based deviation ($q$-merge and $q$-split) possess higher complexity than individual-based deviation. Despite this fact, it is difficult to know in advance which of the deviation models leads to better performance in different applications. Note, that further deviation models, other than the ones mentioned in this lecture note, can be developed, as for example in~\cite{Brandt2015a}.

\begin{figure*}[t]
\centering
\begin{minipage}[c]{7.4cm}
\centering
\includegraphics[width=5.3cm,clip]{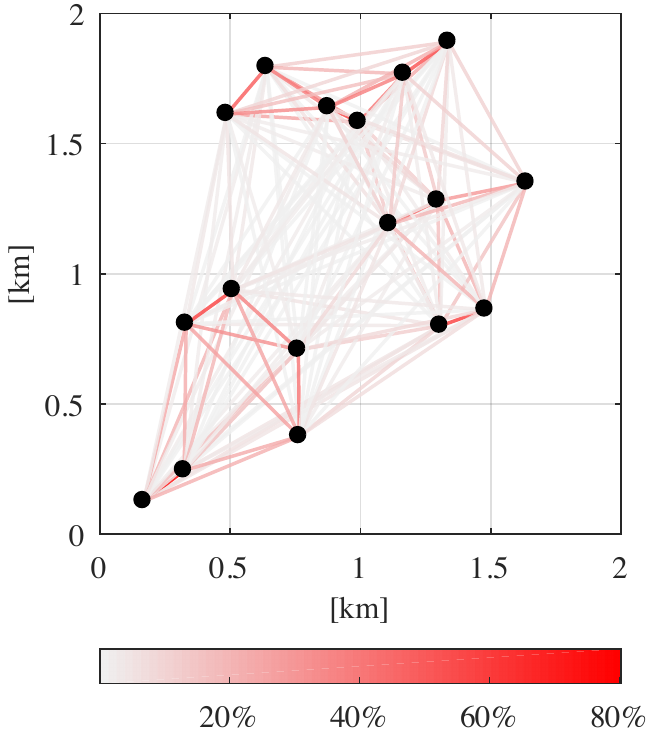}%
\subcaption{\label{fig:topology} Deployment of transmitters. Line intensity between transmitters is the frequency of cooperation.}
\end{minipage}
\hspace{.4cm}
\begin{minipage}[c]{9.5cm}
\centering
\includegraphics[height=5.21cm,clip]{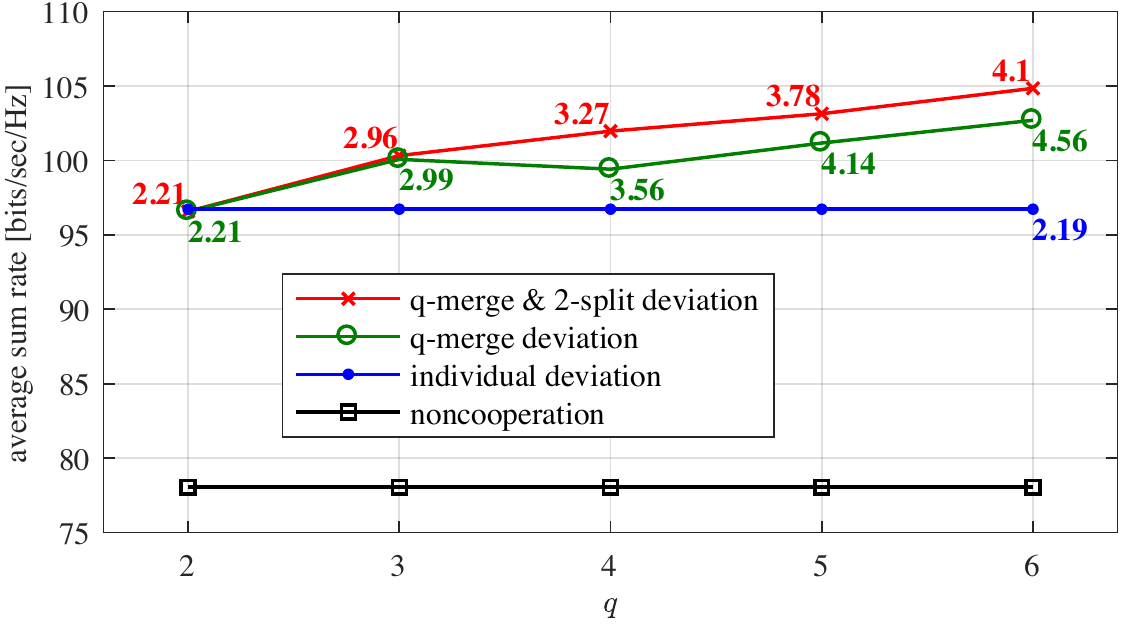}
\subcaption{\label{fig:evaluation} Performance of different coalition formation models depending on parameter $q$ for the setup in \figurename~\ref{fig:topology}. The number at the curves corresponds to the average number of MISO links which cooperate.}
\end{minipage}
\vspace{-0.1cm}
\caption[]{\label{fig:simulation} Coalition formation in the example MISO IC.\footnotemark} \vspace{-0.6cm}
\end{figure*}

\footnotetext{The Matlab simulation code is available online at https://github.com/rami-mochaourab/coalition-formation}

We simulate coalition formation in the considered MISO IC example with $17$ links. The deployment of the transmitters, each with $t = 8$ antennas, is shown in \figurename~\ref{fig:topology}. The receivers' locations are uniformly distributed within $200$ meters from their associated transmitters. We assume Rayleigh small-scale fading, $46$ dBm transmission power at the transmitters, and the following parameters for the setup \cite{Brandt2015a}: \smallskip

\noindent
\begin{center}
\footnotesize
\begin{tabular}{|l|l||l|l||l|l|}
\hline
Path loss & $15.3 + 3.76$ $\log_{10}$ (distance [meters]) & Receiver noise figure & $9$ dB & Noise PSD & $-174$ dBm/Hz \\
Shadow fading & Log-normal i.i.d. with standard deviation $8$ dB & Carrier frequency & $2$ GHz & Bandwidth & $10$ MHz\\
\hline
\end{tabular}
\end{center}
\smallskip

For this setup, the average performance (over $10^3$ realizations) of stable coalition structures reached through coalition formation (Algorithm~\ref{alg:coalition_formation}) with different deviation models and associated preference relations is shown in \figurename~\ref{fig:evaluation}. Here, we always initialize Algorithm~\ref{alg:coalition_formation} with singleton coalitions.

Coalition formation with ``$q$-merge \& $2$-split'' alternates between $q$-merge and $2$-split operations. In order to ensure convergence of this algorithm, as in the individual deviation model, we do not allow a set of coalitions to merge to a coalition which has previously existed. It can be observed that by combining $2$-split with $q$-merge, we allow further dynamics in coalition formation which leads to better performance compared to only $q$-merge deviation. For larger values of $q$, the average coalition sizes of both group-based deviation models increase due to the possibility of merging larger number of coalitions. This, in most cases, leads to improvement in average performance, however at the cost of higher complexity (\figurename~\ref{fig:complexity}). Compared to group-based deviation, individual deviation leads to smaller average coalition sizes and inferior performance, as is also observed in \cite{Mochaourab2014} for a similar MISO IC setting.

In \figurename~\ref{fig:topology}, the line intensity between the transmitters shows the frequency (in percent) of cooperation between any two MISO links when using ``$4$-merge \& $2$-split''. Clearly, it is beneficial to form coalitions between links which are close to each other in order to benefit most from interference management.

\bibliographystyle{IEEEtran}
\bibliography{IEEEabrv,references}

\end{document}